# Thermo-electron accumulation in light and heavy water during MHz-burst laser ablation

Denys Moskal*, Jiří Martan, Vladislav Lang, Milan Honner

New Technologies - Research Centre, University of West Bohemia, Univerzitní 2732/8, 301 00 Pilsen, Czech Republic

* Corresponding author. Tel.: +420-377-634-724; *E-mail address:* moskal@ntc.zcu.cz

**Abstract**

Laser-induced water ablation triggers various physical effects, including atom ionization, optical breakdown of the liquid, phase explosion, cavitation, and shockwave propagation. These effects can be further amplified in heavy water by deuterium–deuterium fusion reactions, which require extremely high energy levels. Laser pulses can be grouped in bursts to achieve the necessary energy within the ablation plasma plume. This study aims to compare the ablation plasma glow and thermal effects in light and heavy water under both single-pulse and burst-mode ultrashort laser irradiation. Notably, this research introduces the novel application of burst laser ablation in heavy water for the first time. The ablation was conducted beneath the water surface along a circular, laser-scanned trajectory, with two distinct ablation regimes: burst mode and single-pulse mode, utilizing lenses with varying focal lengths and different pulse durations. Absorption processes and plasma glow were monitored using visible and infrared detectors, a fast silicon detector, and a thermocouple.

The study revealed that the burst regime in heavy water produced the most intense plasma glow when 1 ps laser pulses were used, with shorter pulses yielding less intense glow and the longest pulses yielding the least. Surprisingly, plasma glow at a lower initial power density of $2.6 \cdot 10^{13}$ W/cm$^2$ was four times higher than at a higher power density of $8 \cdot 10^{13}$ W/cm$^2$. These findings were compared with existing theories on plasma formation in water by ultrashort laser pulses. The observed increase in pulse-to-pulse plasma glow in burst mode was attributed to thermo-electron accumulation effects. The density of excited and hydrated electrons was calculated using both strong-field ionization and avalanche ionization models. Additionally, the influence of pulse parity on burst ablation glow in heavy water was discussed.

## 1. Introduction

Optical breakdown in opaque media under intense laser irradiation is closely associated with a rapid increase in light absorption and forming a dense plasma plume. The absorbed laser energy is efficiently transferred into this plasma, resulting in significant energy redistribution within the medium. During this process, the density of laser-excited free electrons can reach magnitudes on the order of $10^{20}$ cm$^{-3}$, triggering multiple excitation mechanisms. These mechanisms initiation occurs when the laser-stimulated excitation of free electrons surpasses a threshold value up to critical electron density $\sim 10^{21}$ cm$^{-3}$, when the plasma frequency becomes equal to laser frequency [1]–[3].

Applying the ultrashort laser pulses to water, whether light or heavy, offers a unique advantage in achieving the high value of free-electron density through the self-focusing of the laser beam. As a result, the threshold energy required for optical breakdown can be shallow, in the range of $1 - 2$ μJ [4]. This low threshold is particularly beneficial in enabling precise control over the laser-induced effects.

Employing the grouped ultrashort laser pulses in a burst mode, the ablation process in water is significantly enhanced. This enhancement is due to the increased absorption of laser energy during the burst, which is attributed to physical modifications within the irradiated volume between pulses [5]. These modifications lead to cumulative effects, reducing the ablation threshold per pulse in comparison to single-pulse irradiation [6], [7]. Various physical effects are activated during the interaction of laser pulses with water, including localized overheating and the saturation of the irradiated volume with hydrated electrons [8]–[11]. These effects are crucial in improving the efficiency and precision of underwater laser ablation, making it a promising technique for forming specific nanostructures, such as colloidal nanoparticles or hierarchical micro/nanostructures [5], [12], [13].

In addition to nanostructure formation, another significant application of ultrashort laser pulses in burst mode is the irradiation of heavy water (D$_2$O) to drive deuterium-deuterium (D–D) nuclear fusion reactions [14]–[16]. The burst mode facilitates the achievement of the up-critical effects necessary for efficient fusion reactions. The cumulative effects observed in the burst regime can potentially enhance neutron yield (DD-neutron) during such fusion experiments [5].

In this study, we investigate the thermal effects and plasma glow associated with the laser ablation of both light water (H$_2$O) and heavy water (D$_2$O) in both burst and single-pulse regimes. As far as we are aware, this research is the first to explore burst laser ablation in heavy water within the published literature. In our experiments, we focused on detecting plasma glow intensity, observing thermal radiation, and measuring temperature changes during laser ablation. These results were compared with predictions from a first-order model of optical breakdown induced by ultrashort laser pulses. The theoretical model employed in this study incorporates multiphoton, tunneling, and avalanche ionization processes, along with an assessment of heat accumulation and the concentration of hydrated electrons within the irradiated volume. These insights contribute to a deeper understanding of the mechanisms underlying laser ablation in water and lay the groundwork for future applications in both material processing and nuclear fusion research.

## 2. Experimental methods

The laser used in this study was MONACO-1035-40-40 with variable pulse duration from 240 fs up to 10 ps, wavelength 1035 nm, with 1 MHz frequency of laser single pulses and burst regime with 2-5 pulses with 20 ns delay between pulses. The pulse energy was 40 µJ. The laser beam delivery system contains ScanLab galvanometer scanner excelliSCAN-14 with a short focus lens (100 mm and 23 µm spot size) and long focus lens (255 mm and 40 µm spot size). The distilled light water and heavy water (deuterising ≥ 99,9 %) were irradiated along a horizontal ring trajectory with 12 mm diameter and 8 mm beneath the water surface for 30 seconds. The laser irradiated water glow recording was repeated at least three times (n = 3) with intervals 10 to 15 minutes (for water temperature decreasing to the initial level). The masses of the light and heavy water were 20 g and 22.2 g respectively. Table 1 presents the full list of applied parameters. The minimal power density of laser pulses at 10 ps pulse duration is equal to $6.4 \cdot 10^{11}$ W/cm$^2$, which is close to the optical breakdown threshold of water. The maximum achieved power density is about $8 \cdot 10^{13}$ W/cm$^2$ which is one hundred times higher than the optical breakdown threshold for the ultrashort laser pulses [4], [17].

Table 1. Parameters of laser scanning of the 12 mm ring beneath the water surface with short and long focus objectives

|  | Single pulses | Burst regime |
|---|---|---|
| Frequency (kHz) | 1000 | 500, 330, 250, 200 |
| Pulses - interval (ns) | 1-1000 | 2-20, 3-20, 4-20, 5-20 |
| Scanning speed (m/s) | 14 (F = 100 mm / angle 7°) 20 (F = 255 mm / angle 2°) | |
| Spot's distance (µm) | 14, 28, 42.4, 56, 70 (F = 100 mm / angle 7°) 20, 40, 60.6, 80, 100 (F = 255 mm / angle 2°) | |
| Pulse duration (fs) | 240, 1000, 10000 | 240, 1000, 10000 |

The experimental setup is illustrated in Fig. 1, featuring several detectors and cameras. The intensity of plasma glow during water ablation was measured using a Thorlabs Si UV-NIR diode (0.3 – 1.1 µm) paired with a LeCroy WavePro 404HD oscilloscope (20 GS/s). The thermal flow from the scanned ring was analyzed using an Optris Xi 400 IR camera (8 – 14 µm, 18° optics), capturing the difference in maximum temperature fields between the upper and lower parts of the scanned volume (Fig. 1, *b*). An optical camera (Canon EOS 70D with a ø 62 mm lens and an L7+1 laser radiation filter) was used to assess plasma glow intensity by averaging the entire ablation ring area (Fig. 1, *c*). Temperature changes were monitored using a K-type thermocouple connected to an Agilent-34970A data acquisition module equipped with a 34902A ADC card (Fig. 1, *a*).

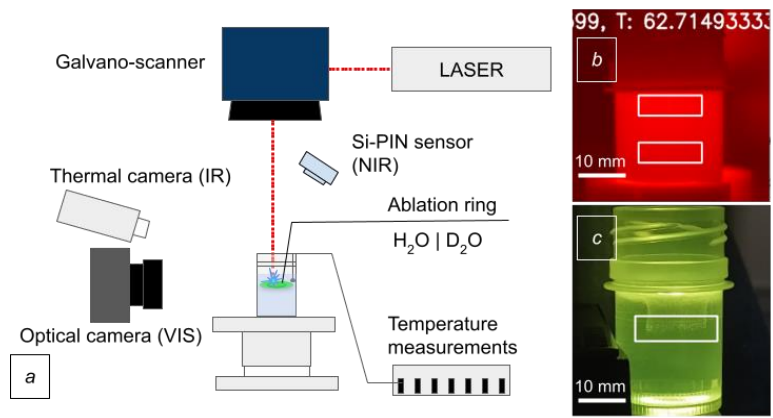

Fig. 1. Laser water ablation experiment: *a* - experimental setup; *b* - IR-camera image with two detected areas (measurement of $T_{Upper}$ and $T_{Bottom}$); *c* - image of the vessel with ablation ring in VIS spectrum

In these experiments, we conducted water ablation directly, free from solid walls or suspended particles, which ensured that the thermal effects observed were exclusively due to the self-focusing of the laser beam in water and the excitation of electrons in a strong electromagnetic field [4]. The water was contained in an open polypropylene container set on a mirror-polished metallic plate. Polypropylene offers about 80% transmittance in the VIS-NIR spectrum [18], [19].

## 3. Experimental results
### 3.1. Ablation glow and IR measurements

A key effect of the burst regime was observed in a series of experiments with a focusing length of $F = 100$ mm (spot size 23 μm). Single-pulse ablation showed minimal glow and did not result in significant cavitation bubble spreading (Fig. 2a). In contrast, the burst regime produced increased ablation glow and cavitation bubbles spread up to 6 mm for 1 ps pulses (compare Fig. 2, *a* with Fig. 2, *b*). It is important to note that the laser power and power density were identical for both the single-pulse and burst regimes. Thus, the intense plasma glow was only observed when the laser pulses were grouped into bursts, despite having the same power density (Fig. 2, *c*).

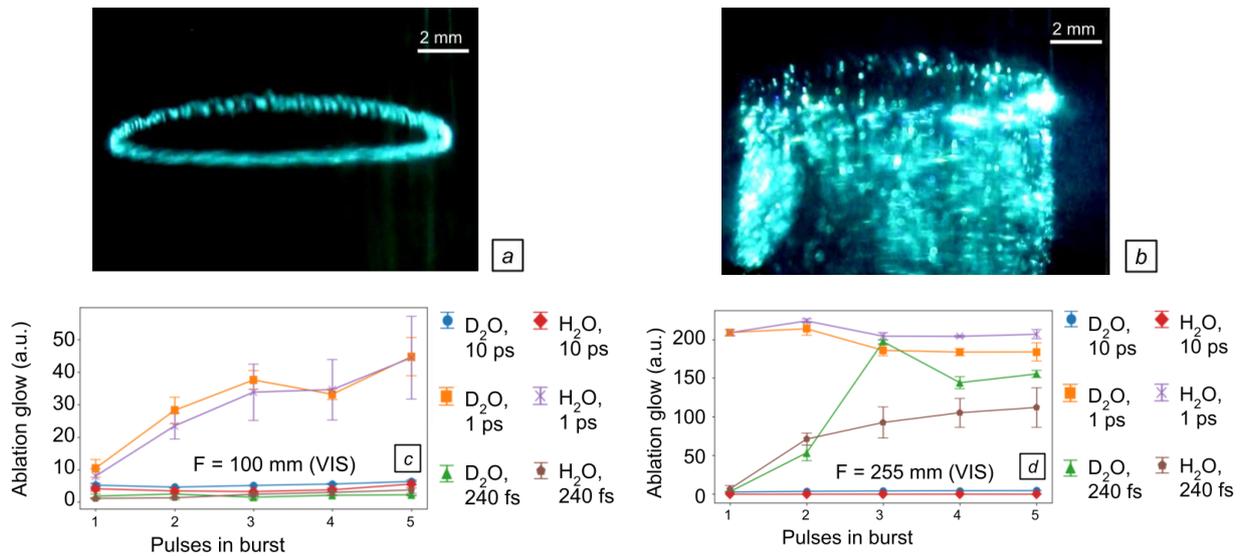

Fig. 2. Experimental results of laser ablation beneath water surface: *a* - scanning ring in heavy water at 1 ps pulse duration and single pulse regime ($F = 100$ mm); *b* - scanning ring in heavy water at 1 ps pulse duration and burst regime with 5 pulses ($F = 100$ mm); *c* - ablation glow in optical spectrum for different pulse number and three pulse durations: 240 fs, 1 ps, 10 ps ($F = 100$ mm); *d* - ablation glow in optical spectrum for different pulse number and three pulse durations: 240 fs, 1 ps, 10 ps ($F = 255$ mm)

In the configuration with a longer focal length ($F = 255$ mm), cavitation bubble spreading and intense ablation glow were observed even with single 1 ps laser pulses (Fig. 2. *d*). Despite the initial laser power density being three times lower in this case ($6.4 \cdot 10^{12}$ W/cm$^2$) compared to the $F = 100$ mm configuration ($1.9 \cdot 10^{13}$ W/cm$^2$), the plasma glow radiation was paradoxically five times higher for the longer focal length experiments. Additionally, in the long-focus configuration, the burst accumulation effect was noted not only for 1 ps pulses but also for 240 fs pulses (with a power density of $2.6 \cdot 10^{13}$ W/cm$^2$, compared to $8 \cdot 10^{13}$ W/cm$^2$ for the same pulse duration at $F = 100$ mm). The longest laser pulses did not produce intense ablation glow in either the long-focus or short-focus experiments. This indicates that the burst regime in water is more efficient for shorter laser pulses. The decrease in plasma formation threshold is likely due to a stronger self-focusing effect of the laser beam with the longer focal length and narrower focusing angle [20], [21].
.

The impact of the burst regime on laser beam reflection was discovered in experiments with the NIR-sensor. Here the detected maximal signal is mainly scattering of the laser beam light. In the case of 1 ps laser pulses and short focus lens ($F = 100$ mm) the detected NIR radiation becomes more than two times higher in the burst regime (Fig. 3, *a*). The interaction of laser pulses in the focal volume during the early onset of self-focusing effects is less intense than the more pronounced self-focusing observed in burst mode [17], [22]. Such an effect of NIR-signal increasing in burst regime was nearly two times lower in experiments with the long focal length lens ($F = 255$ mm, Fig. 3. *b*). This variation between burst regimes in short and long-focal length experiments can be explained by the enhanced efficiency of light absorption in the laser-ablated area with a 255 mm focal length. In other words, the bigger part of light is absorbed in water during ablation with a narrow laser beam self-focusing process.

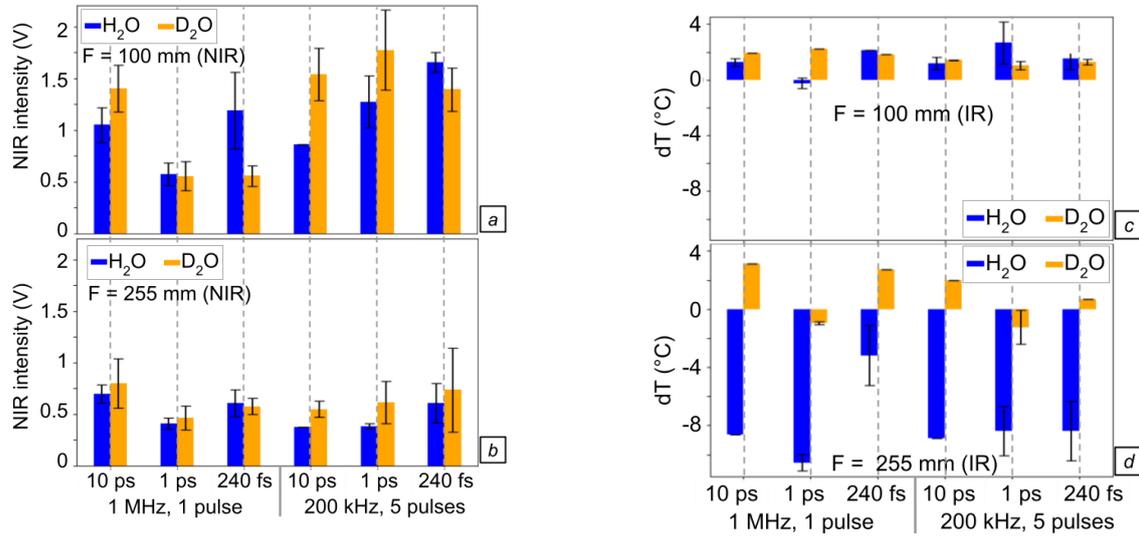

Fig. 3. Measurements in infrared spectrum during laser ablation beneath water surface: *a* - detection in NIR spectrum ($F = 100$ mm); *b* - detection in NIR spectrum ($F = 255$ mm); *c* - temperature difference between bottom and upper area ($T_{Bottom} - T_{Upper}$), detected in IR spectrum ($F = 100$ mm); *d* - temperature difference between bottom and upper area ($T_{Bottom} - T_{Upper}$), detected in IR spectrum ($F = 255$ mm).

Higher absorption in experiments with longer focus lenses was confirmed by measurements with the IR-thermal camera (8 – 14 μm), where the biggest energy collects in the scanning ring plane. In comparison, in the experiments with the short focus lens (with lower ablation effectivity) the scanning ring plane is colder than the area under the ablation ring (Fig. 3. *c*). However, the temperature of the ablation ring is higher in experiments with a long focal length compared to the temperature changes observed in the area below the ablation ring (Fig. 3, *d*). Notably, heavy water exhibits lower sensitivity to the focusing angle, as illustrated by the data in Fig. 3, *c* and Fig. 3, *d* (orange bars).

### 3.2. Thermocouple measurements

The previous paragraph demonstrated that a narrow laser beam with a focal length of 255 mm and a focusing angle of 2° produces a higher plasma glow. The wide angle laser beam ($F = 100$ mm, angle 7°) brings weaker excitation of electrons. Nevertheless, in both scenarios, the laser beam continues to absorb energy along its entire trajectory, extending beyond the focal volume. Such a non-ablation absorption of the laser beam energy leads to volumetric heating of water as it was detected by thermocouple (Fig. 4). These measurements with thermocouple indicate how the laser ablation processes and self-focusing affects energy absorption in water volume. For the short focus distance ($F = 100$ mm), light water exhibits lower thermal accumulation compared to heavy water (Fig. 4, *a*). This observation is consistent with the higher optical density (absorption) of heavy water [23]. This suggests that in experiments with a short focus distance, self-focusing and nonlinear processes are suppressed, leading to a greater portion of the energy being absorbed linearly.

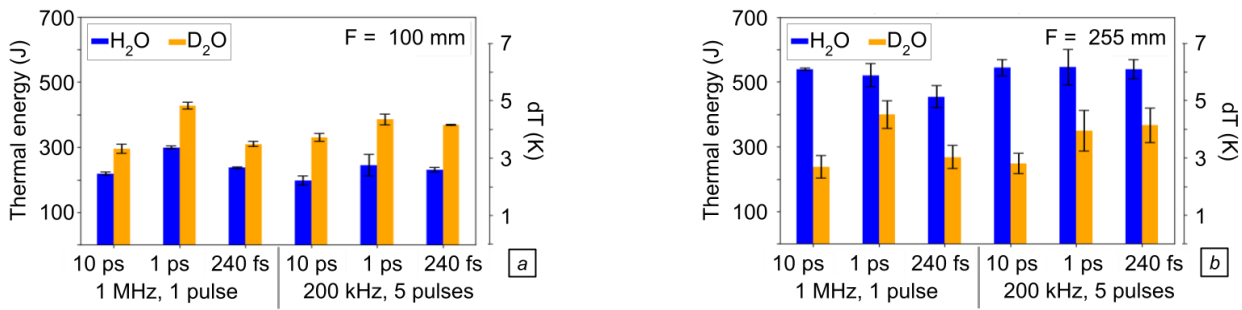

Fig. 4. Evaluation of heat balance and temperature changes measured with thermocouple: *a* - in experiments with 100 mm focus length; *b* - in experiments with 255 mm focus length.

The significant role of non-linear processes and the influence of self-focusing underlies higher heat accumulation of water in experiments with long focus distances (Fig. 4, *b*). The light water in this case has absorbed nearly two times higher energy in both regimes, - single laser pulses or bursts. On the other hand, the experiments with heavy water indicate lower affection of the light absorption by nonlinear processes and energy accumulation is near the same for the bot experiments - with short focus and long focus distances (compare heavy water results on Fig. 4, *a* and Fig. 4, *b*).

Another notable feature is the nearly equal thermal energy accumulation observed for both single distanced laser pulses and burst laser pulses in the same type of water. In contrast, a significant effect of the burst regime was detected in plasma glow measurements (Fig. 2, *c* and *d*), where an increase in the number of pulses directly led to a rise in plasma glow intensity. This discrepancy between thermal measurements and plasma glow measurements suggests that the majority of the absorbed energy remains within the water volume, while only a small fraction of the laser pulse energy contributes to the ablation glow process.

## 4. Theoretical modelling of water excitation
### 4.1. Laser pulse induced ionization

Optical breakdown refers to a significant increase in light absorption in opaque media, driven by laser-induced ionization of molecules [1], [4], [24], [25]. This intense light absorption within the focal volume leads to the formation of hot plasma and, consequently, the ablation of the irradiated material. In the case of water, the laser-induced ionization can be likened to the excitation of electrons in a wide-band gap semiconductor, where dopant levels are created by initial traps and hydrated electron states [3], [26]. The band gap in water is equal to $E_g = 9.5$ eV, while the interband trap level associated with preexisting traps and excitation via hydrated electrons is $E_{ini} = 6.4$ eV [3]. According to Keldysh's theory, optical breakdown in such a wide-band gap medium occurs due to the increased probability of ionizations in a strong electromagnetic field[2]. The total electron excitation rate can be described as a combination of strong field ionization (SFI) and avalanche ionization (AI) [1], [3], [4]:

$$\left(\frac{d\rho_{ex}}{dt}\right) = \left(\frac{d\rho_{SFI}}{dt}\right) + \left(\frac{d\rho_{AI}}{dt}\right) \cdot \rho_c \tag{1}$$

where $\rho_{ex}$ - is the density of laser-excited electrons, $\rho_{SFI}$ - the density of electrons excited by SFI, $\rho_{AI}$ - is the density of free electrons generated by AI and $\rho_c$ - density of free electrons in the conduction band. Here it is needed to underline that the first term in Eq. 1 includes two processes of water ionization in the strong electromagnet field of laser-direct multiphoton ionization and ionization on initial centers, like hydrated electrons, ions and initial impurities, as it was detailed described by Linz et al. [3]:

$$\left(\frac{d\rho_{SFI}}{dt}\right) = \left(\frac{d\rho_{\text{ini}}}{dt}\right) + \left(\frac{d\rho_{\text{Egap}}}{dt}\right) \tag{2}$$

with

$$\left(\frac{d\rho_{\text{ini}}}{dt}\right) = \eta_{SFI}(\text{E}_{ini}) \times \left(1 - \frac{\rho_{ini}}{\rho_{ini,max}}\right) \tag{3}$$

$$\frac{d\rho_{\text{Egap}}}{dt} = \eta_{SFI}(\tilde{\Delta}) \tag{4}$$

here $\eta_{SFI}(\text{E}_{ini})$ - excitation rate which is defined by Keldysh theory [1], [2], $\rho_{ini}$ - density of electrons excited through initiation channel and $\rho_{ini,max}$ - finite capacity of the initiation channel. The full rate equation for electrons in the conduction band should account for both recombination and diffusion processes:

$$\left(\frac{d\rho_c}{dt}\right) = \left(\frac{d\rho_{ex}}{dt}\right) - n_{diff} \cdot \rho_c - n_{rec} \cdot \rho_c^2 \tag{5}$$

where $n_{diff}$ - diffusion rate of electrons and $n_{rec}$ - recombination rate [4]. The resulting density of free electrons, which was exited during laser pulse in water was calculated numerically according to theoretical derivations presented by Kennedy in his basic work [1] and then this theory was developed in several latter works by A. Vogel, N. Linz, and X. Liang [3], [4], [25], [27].

The main input parameters of the laser are pulse duration, wavelength, and power density. In the calculation model, the focal volume was described as an area with randomly distributed points with different power densities (Fig. 5). The resulting maximal power density in the focal center becomes smaller compared to the irradiation of the flat surface of solids. However, the affection of the irradiated volume by the laser beam leads to modification of optical properties of the media with laser beam self-focusing effect [20], [24], [25]. Self-focusing again significantly increases the power density at the focal plane. This effect of self-focusing was incorporated by reducing the laser beam diameter, which is constrained by the critical density of free electrons $\rho_{cr}$. This critical regime of laser water irradiation was associated with the maximum plasma glow experimentally observed. This means that the minimal diameter of the self-focused laser beam, constrained by the critical electron density ($\rho_{cr}$), corresponds to the laser settings at which the maximum plasma glow was observed.

The logical scheme of self-focusing diameter evaluation can be written as a chain: maximal plasma glow → critical electron density $\rho_{cr}$ → self-focused plasma diameter. In our experiments, the maximum plasma glow, measuring 208.9 a.u. from the ablation ring, was observed with 1 ps laser pulses (see Fig. 2, *d*). The critical electron density for wavelength $\lambda = 1064$ nm is equal to $\rho_{cr} \approx 0.98 \cdot 10^{21}$ cm$^{-3}$ (and it was taken the same for experimentally used wavelength $\lambda = 1035$ nm ) [4]. For a 1 ps laser pulse duration, the theoretical prediction for achieving critical electron density indicates that the laser beam's self-focusing diameter should be 15.58 μm.

The breakdown threshold of excited electrons density $\rho_{th} = 1.8 \cdot 10^{20}$ cm$^{-3}$ was predefined in the temperature balance equation [3], [28]:

$$\rho_{th} = \frac{p_0 C_p \Delta T_{th}}{\left(\frac{9}{4}\right)\tilde{\Delta}}, \tag{6}$$

where $p_0$ - water density, $C_p$ heat capacity $\Delta T_{th}$ - threshold temperature rise around 147.7 K [28] and $\tilde{\Delta}$ - denotes effective ionization potential in a strong electromagnetic field (determined by Keldysh theory [1], [2]). The lower threshold of $1.8 \cdot 10^{20}$ cm$^{-3}$ was used as the initial parameter for evaluating the self-focusing diameter, which corresponds to a measured plasma glow of 6 a.u. (when the accumulation effect in a burst is most pronounced). This bonding between measured plasma glow and excited electrons density in the up-threshold range of electron density was used as a fitting function (in cm$^{-3}$):

$$\rho_c = G \cdot 3.96 \cdot 10^{18} + 1.52 \cdot 10^{20} \tag{7}$$

where $\rho_c$ is the estimated free electron density and $G$ the measured plasma glow in the visual spectrum. By fitting this function with measured plasma glow intensities, we can evaluate the electron density and, consequently, predict the theoretical diameter of the self-focusing laser beam, as well as the temperature and hydrated electron density in the irradiated volume. Table 2 represents the predicted diameters of the self-focused laser beams according to the reached plasma glow intensity. In Table 2, the self-focused diameter was bonded with plasma glow after the 1x laser pulse.

Table 2. Diameters of the self-focused laser beams with corresponding laser pulse duration and focal length of scanning system in burst regime

| Focusing length (mm) | Pulse duration (fs) | 1x plasma glow $H_2O$ \| $D_2O$ (a. u.) | 5x burst plasma glow $H_2O$ \| $D_2O$ (a. u.) | Electron density 1x $H_2O$ \| $D_2O$ ($cm^{-3}$) | Electron density 5x $H_2O$ \| $D_2O$ ($cm^{-3}$) | Self-focusing $H_2O$ \| $D_2O$ (μm) |
|---|---|---|---|---|---|---|
| 255 | 240 | 6.4 \| 2.9 | 112.1 \| 154.7 | $1.78 \cdot 10^{20}$ \| $1.64 \cdot 10^{20}$ | $5.96 \cdot 10^{20}$ \| $7.65 \cdot 10^{20}$ | 18.58 \| 18.34 |
|  | 1000 | 208.9 \| 209.2 | 206.9 \| 184.1 | $9.80 \cdot 10^{20}$ \| $9.81 \cdot 10^{20}$ | $9.72 \cdot 10^{20}$ \| $8.81 \cdot 10^{20}$ | 15.58 \| 15.36 |
|  | 10000 | 1.0 \| 2.6 | 4.4 \| 4.4 | $1.56 \cdot 10^{20}$ \| $1.63 \cdot 10^{20}$ | $1.70 \cdot 10^{20}$ \| $1.70 \cdot 10^{20}$ | 11.23 \| 11.06 |
| 100 | 240 | 1.1 \| 1.8 | 3.68 \| 2.22 | $1.57 \cdot 10^{20}$ \| $1.59 \cdot 10^{20}$ | $1.67 \cdot 10^{20}$ \| $1.61 \cdot 10^{20}$ | 18.62 \| 18.36 |
|  | 1000 | 8.0 \| 10.4 | 44.4 \| 44.7 | $1.84 \cdot 10^{20}$ \| $1.93 \cdot 10^{20}$ | $3.28 \cdot 10^{20}$ \| $3.29 \cdot 10^{20}$ | 15.91 \| 15.68 |
|  | 10000 | 4.1 \| 5.2 | 5.52 \| 6.32 | $1.69 \cdot 10^{20}$ \| $1.73 \cdot 10^{20}$ | $1.74 \cdot 10^{20}$ \| $1.77 \cdot 10^{20}$ | 11.22 \| 11.07 |

As shown in Table 2, even a small change in laser spot diameter leads to significant variations in electron density. For instance, with a 240 fs laser pulse in water, the estimated spot diameter for sub-threshold laser irradiation is 18.62 μm, while for above-threshold irradiation, the predicted diameter is 18.58 μm. The difference between these sub-threshold and above-threshold self-focusing diameters is noticeable only in the first decimal place. This sensitivity in the calculations can be attributed to the exponential dependence of electron density on intensity changes. In general, this relationship can be expressed as [1], [29]:

$$\rho_c = \Delta t \cdot f \cdot I_0^k \qquad (8)$$

where $\Delta t$ represents a fraction of the pulse duration ($\Delta t = \tau_p/10$), corresponding to the most productive time for free electron generation, $f$ is a function that describes the number of ionizations per unit volume per unit time, $I_0$ is the peak laser irradiance (W/m²), and $k$ is the minimum number of collisional absorption events, which can also be understood as the wavelength-to-effective-band-gap ratio. For the used wavelength the power exponent $k$ can be as high as 12-13, increasing further with higher light intensities [27]. This high value of the power exponent is the primary reason for the extreme sensitivity of predicted free electron density to small changes in spot diameter.

The self-focusing diameter was defined from the laser-irradiated volume of water with an ellipsoidal form with Gaussian energy distribution. Here the minor axis corresponds to the diameter of the self-focused laser beam and the major axis was taken equal to the Rayleigh range [25]:

$$d_s = 2 \cdot r_s \qquad (9)$$

and

$$l = \frac{z_R}{2} = \pi d_s^2 \left(\lambda / 2n\right) \qquad (10)$$

where $d_s$ - diameter of the self-focused laser beam in the focal plane, $r_s$ - radii of self-focusing, $l$ - major axis length, $z_R$ - Rayleigh range, - wavelength, $n$ - refraction index. The intensity distribution $I$ in the ellipsoidal focal volume was described by a Gaussian-like function [4] with normalization over the volume:

$$I(r,z) = I_0 exp\left[-2\left(\frac{r^2}{r_{self}^2} + \frac{z^2}{l}\right)\right] \qquad (11)$$

where $I_0$ - the normalized central point laser intensity, $r$ - the distance from central point in the focal plane, and $z$ - in-the-beam distance from the central focusing point. The space-time distribution of intensity essentially reflects the distribution of free electrons within the focal volume. The power density of the laser pulse is defined by probability density functions (PDFs), each of which is normalized by the full energy of the laser pulse in the randomly distributed points of the irradiated volume (Fig. 5). The ablation volume can be divided by three conditional zones, similar to dividing excited electrons by three levels: sub-threshold, threshold and critical. In the case of near threshold water irradiation, the surface of the ablation core is several times smaller in comparison to the critical regime, and as result the inner plasma should have several times higher pressure for explosion of the core.

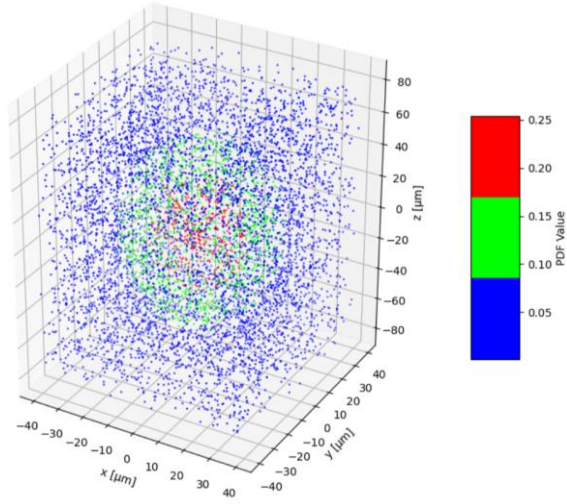

Fig. 5. Gaussian-like 3D-distribution of power density in the local volume of the self-focused vertical beam as probability density function (PDF).

The time-space distribution of laser power density plays a crucial role in the calculation of the avalanche electron density ($\rho_{AI}$ in Eq. 1). The avalanche mechanism is probabilistic and requires at least one "seed" electron, generated during photoionization, which is proportional to a power function of the laser beam intensity. [1], [4], [27]. The probability of avalanche ionization is described by the Heaviside function:

$$\theta_{AI} = \begin{matrix} 1 \ at \ \rho_c(I^k)dV \geq 0.5 \\ 0 \ at \ \rho_c(I^k)dV < 0.5 \end{matrix} \quad (12)$$

where $dV$ represents the discrete volume in the irradiated region around a point in Fig. 5 [1], [4]. The resulting Heaviside function AI is then used as a logical key for inclusion in Eq. 1 the avalanche mechanism of ionization. This distribution of excited electrons and the intensity of the self-focused laser beam were then used to evaluate the dynamics of laser energy absorption and transfer, including the generation and accumulation of excited and hydrated electrons, as well as heat accumulation from pulse to pulse in a burst mode.

### 4.2. Burst regime and hydration of free electrons

The biggest part of the excited electrons in the conduction band will be solvated during hydration time $t_h = 300$ fs [9], [10]. In the hydrated state, electrons are surrounded by three, four, six, or more molecules of water and they will be fixed in this stable state during a long time interval, up to several microseconds [30]–[33]. The lifetime constant of hydrated electrons was determined to be $t_g = 300$ ns [10]. This combination of a short hydration time and a long lifetime facilitates the accumulation of hydrated electrons between laser pulses in a burst. The density of hydrated electrons was defined by the reduction in the number of free electrons, which is attributed to their trapping within water clusters:

$$\rho_h = \rho_c(t) \cdot \left(1 - e^{-t/t_h}\right) \quad (13)$$

This equation was included in numerical calculations, where the conduction electron density is a function of ionization rate and recombination (see Eq. 3 and 5). The same principle of electron density dynamic was included for the evaluation of transferring electrons into the ground state, but with a longer grounding time constant $t_g$ in relation to the density of hydrated electrons.

The diffusion length of the hydrated electrons during the interval between laser pulses can be calculated from the diffusion coefficient and in-burst pulse frequency:

$$L_h = \sqrt{D_h \cdot t_B} \quad (14)$$

where $D_h$ - diffusion coefficient of hydrated electrons equal to $4.9 \cdot 10^{-5}$ cm$^2$s$^{-1}$ and $t_B$ is the in-burst time interval. In our experiments, the burst pulse time interval was equal to $t_B = 20$ ns and it gives the diffusion length $L_h \sim 10$ nm only. The diameter of the focal volume is in the range of tens of micrometers, significantly larger than the evaluated diffusion length. Consequently, hydrated electrons are considered to be localized within the focal volume during the burst [34]. According to Linz et al., hydrated electrons can act as initial centers for the subsequent up-conversion of electrons into the conduction band, with an energy level of approximately $E_{ini} \approx E_h = 6.4$ eV [3]. For evaluation of the transferring rate to the ground state of the hydrated electrons the lifetime $\tau_{e^-} = 300$ ns was used [10]. Such a relatively long lifetime of hydrated electrons leads to an increase of initial centers density with energy level $E_{ini}$ in water band gap during laser pulses burst:

$$\rho_{ini,max}^{i+1} = \rho_h^i + \rho_{ini,max} \quad (15)$$

where $\rho_{ini,max}^{i+1}$ - density of initial traps for $(i + 1)_{th}$ laser pulse in a burst, $\rho_h^i$ - density of the hydrated electrons produced during $i$ laser pulses. The increased density of intraband traps $\rho_{ini,max}^{i+1}$ affects the excitation intensity and is incorporated as a new upper parameter in Eq. 3 instead of the initial value $\rho_{ini,max}$.

The heat accumulation in the burst regime was evaluated as a heaping of thermal energy in the burst from pulse to pulse, similar to hydrated electron accumulation. The temperature rise $\Delta T$ before every next laser pulse is defined by full energy which is involved in ionization and hydration processes [25], [27]:

$$\Delta T = \frac{U}{p_0 \cdot C_p} \quad (16)$$

where U - the full thermal energy involved by ionization and hydration of laser-excited electrons. The thermal energy accumulated in a burst is an integration of recombined electron density energy and a part of energy brought through the hydration state:

$$U \simeq \frac{9\tilde{\Delta}}{4}(\rho_c - \rho_h) + E_h \rho_g \quad (17)$$

where $\rho_g$ - density of electrons transferred at ground level through the hydrated state during burst laser pulse interval. It is need to note, that Eq. 17 was solved numerically with the inheritance of the hydrated electron density $\rho_h$ as initial traps for every next laser pulse of the burst. The theoretical thermal energy absorbed in the ablation volume represents only a portion of the total energy absorbed by the material, as linear absorption of light occurs along the entire beam path through the irradiated medium.

The ablation volume absorbed energy can be compared with full thermal energy absorbed in water. The full path absorbed laser energy can be evaluated by the thermal balance equation from temperature changes, measured on the thermocouple:

$$U_F = \frac{\Delta T_F \cdot C_p \cdot m}{N_F} \quad (18)$$

where $U_F$ - is the full absorbed light energy per pulse, $\Delta T_F$ - temperature changes (detected by the thermocouple), $m$ - full mass of the irradiated water, and $N_F$ - full number of laser pulses during the scanning of the ablation ring. Such heat accumulation is a long-term process and indicates only general heat accumulation. For comparison, the evaluation of the thermal energy accumulation in a short time interval between laser pulses in a burst was done theoretically. During 20 ns intervals between laser pulses in a burst, the thermal diffusion length achieves $L_{heat} \approx 70$ nm. For the full burst ending thermal diffusion length has not increased a lot and becomes $L_{Burst} \approx 280$ nm only. Such a short thermal diffusion length allows to describe the heat accumulation process as a sum [35]–[37]:

$$\Delta T_{Burst} = \frac{1}{4 \cdot p_0 \cdot C_p} \sum_{i=1}^{N_{Burst}} U_i \quad (19)$$

where $i$ - is an index of the laser pulse in the laser burst, $N_{Burst}$ - is the full number of laser pulses in the burst and $U_i$ - is the absorbed energy during the one laser pulse. The evaluation of thermally accumulated energy in light and heavy water took into account differences in physical properties, including water density, heat capacity, molecular mass, and optical absorptivity.

## 5. Modelling results
### 5.1. Comparison of hydrated electrons accumulation with burst laser ablation

*Irradiation with long focus length ($F = 255$ mm)*

It was mentioned that the maximal plasma glow was detected in experiments with 1 ps laser pulses and it keeps on near the same level for all laser pulses number in the laser burst and longer focus distance ($F = 255$ mm) (Fig. 2, *d*). This phenomenon can be explained by the improved filamentation of the laser beam at a smaller focusing angle [20]. The detected maximum plasma glow was used as a reference point between the experimental settings of the laser and the diameter of the self-focusing laser beam. It was taken to be equal to 15.58 μm and 15.36 μm in light and heavy water respectively (Table 2). In this case, the critical free electron density $\rho_{cr}$ was achieved already during the first laser pulse in the burst. It was noticed that between burst laser pulses the bigger part of the excited electrons are accumulated in the ablated volume in the hydrated state due to short hydration time and relatively long lifetime in the hydration state [9], [11]. The first laser pulse theoretically brings the density of the hydrated electrons near $\rho_{h,1x} = 2.2 \cdot 10^{21}$ cm$^{-3}$. This value of the hydrated electron density corresponds to the order of peak magnitude obtained in experimental work [11]. The predicted maximum of the accumulated density of the hydrated electrons at the end of the fifth laser pulse in the burst reached order $10^{22}$ cm$^{-3}$ during irradiation with 1 ps laser pulses.

The hydrated electrons have an energy level near 6.6 eV in the band gap of the water and they can work as initiators for every next laser pulse in the laser burst (Eq. 15). The hydrated electrons accumulation should increase water absorption [8] from pulse-to-pulse during laser burst irradiation. The energy of the ablation plume will be higher and should bring a higher plasma glow with increasing laser pulse quantity in a burst. However, the part of the free electrons brought into the conduction band through the initiation channel is low, more than two orders lower in comparison with the AI channel [3]. Moreover, the laser electrons excitation by laser in water is principally limited by full ionization, and then the higher number of laser pulses is not able to increase the plasma glow intensity by increasing the hydrated electron density (compare Fig. 2, *d* and Fig. 6, *c*). The high density of excited electrons influences the optical properties, and this principle similarly applies to limiting the density of hydrated electrons, leading to a localized reduction in the refractive index [38]. The laser beam self-focusing by opto-thermal process and excitation de-focusing are going in opposite directions [38]. The plasma glow no longer increased with each subsequent laser pulse in the burst, indicating it is nearing saturation (Fig. 2, *d*), as the ionization rate approaches the order of magnitude of the atomic density in water, approximately $\sim 10^{22}$ cm$^{-3}$. In this case the avalanche ionization is reduced by statistical principle of it and the laser beam reflection increases on high density ablation plasma plume.

Every next laser pulse on its path meets the water volume already obstructed after the previous laser pulse and it can additionally suppress the self-focusing mechanism. This is evidenced by the greater thickness of the ablation ring in the burst regime compared to distant pulse scanning (see Fig. 6, *a* and *b*). The higher density of hydrated electrons $\rho_h \sim 10^{22}$ cm$^{-3}$ brings intense transferring of electrons into the ground state (Eq. 13 in relation to hydrated electrons and grounded electrons). Such evaluation gives only a relative understanding of the thermo-electric accumulation in the irradiated volume. It should be noted that the model did not consider the dynamics of hydrated electrons or other physical effects such as cluster interactions, hydration state saturation, and shock wave propagation. Consequently, the real $\rho_h$ in the burst regime is likely to be lower than the predicted values.

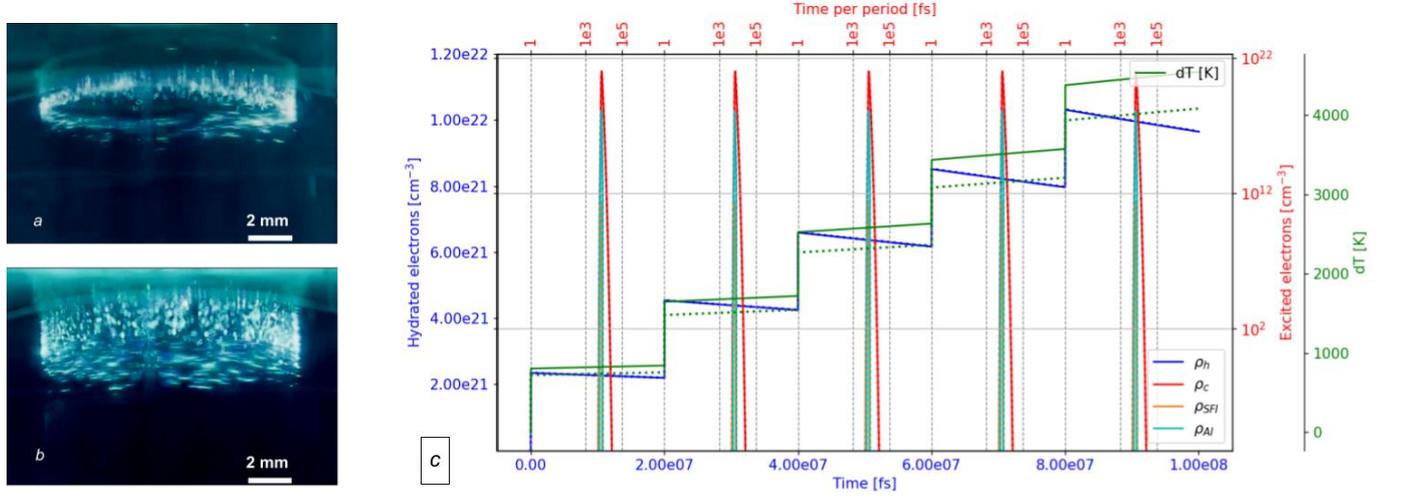

Fig. 6. Single pulses and burst water ablation ($\tau_p = 1$ ps, $F = 255$ mm): *a* - ablation ring in single distant pulses regime; *b* - ablation ring in burst regime with 5x pulses; *c* - theoretical evaluation of excited electron density dynamic, accumulation of hydrated electrons and temperature (solid lines - H$_2$O, dotted lines - D$_2$O).

The influence of pulse duration on self-focusing is confirmed by the data presented in Table 2. The maximal ablation plasma glow was 1.5 - 2 times higher in the case of 1 ps in comparison to 240 fs and 10 ps laser pulses. The self-focusing of the laser beam with 240 fs has the lowest self-focusing 18.58 μm and 18.34 μm for light and heavy water respectively. Such reduced self-focusing effect during the shortest laser pulse can be explained by lower thermal affection of a smaller area during the short laser pulse. In the shortest laser pulses, the duration has not reached the hydration time and the most excited electrons recombine. However, even with a lower accumulation of hydrated electrons, the combination with high power density plays a crucial role in plasma formation during the burst regime. This is evidenced by the increased plasma glow observed in near-threshold water ablation conditions, as demonstrated in the experiments with 240 fs laser pulses and a 255 mm focusing lens shown in Fig. 7. The plasma glow initiated by the distant single laser pulses was tens times lower in comparison to the burst regimes with 5x laser pulses (see Fig. 2, *d*). The accumulation of hydrated electrons, in this case, was defined as $\rho_{h,1} = 2.57 \cdot 10^{20}$ cm$^{-3}$ for the first laser pulse and $\rho_{h,5} = 1.06 \cdot 10^{21}$ cm$^{-3}$ for the last laser pulse in a full burst in light water. In heavy water, the accumulation of hydrated electrons in the burst mode is slightly lower and achieves $\rho_{h,1} = 2.42 \cdot 10^{20}$ cm$^{-3}$ for the first laser pulse and $\rho_{h,5} = 9.93 \cdot 10^{20}$ cm$^{-3}$.

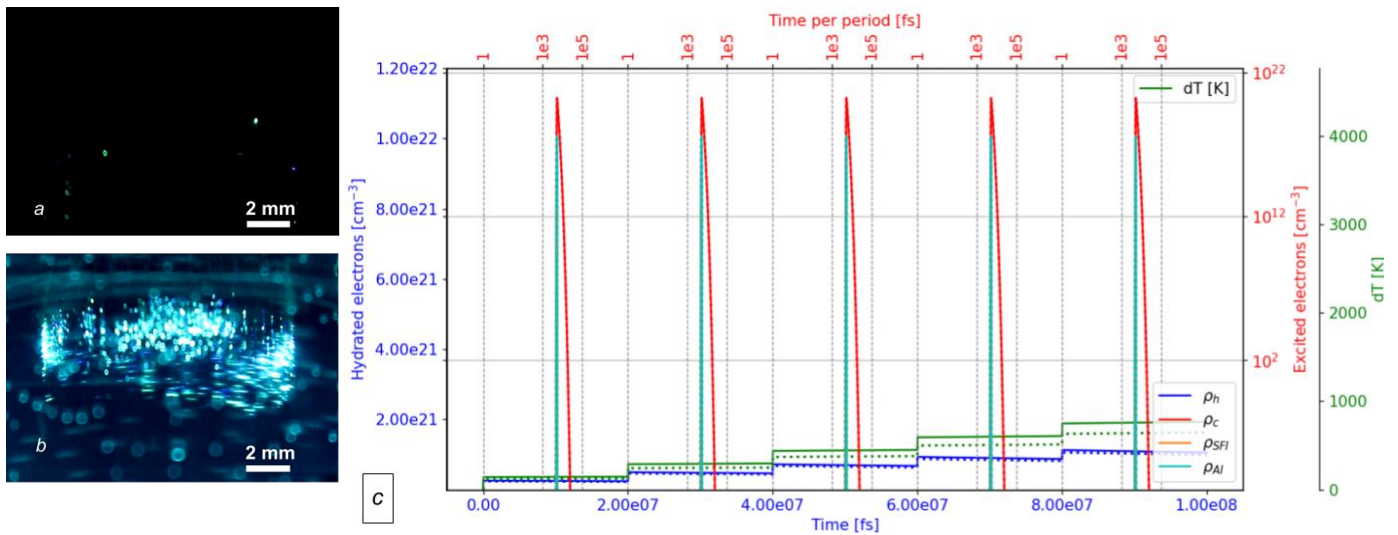

Fig. 7. Single pulses and burst regimes of water ablation ($\tau_p = 240$ fs, $F = 255$ mm): *a* - ablation ring in single distant pulses regime; *b* - ablation ring in burst regime with 5x pulses; *c* - theoretical evaluation of excited electron density dynamic and accumulation of the hydrated electrons (solid lines - $H_2O$, dotted lines - $D_2O$)

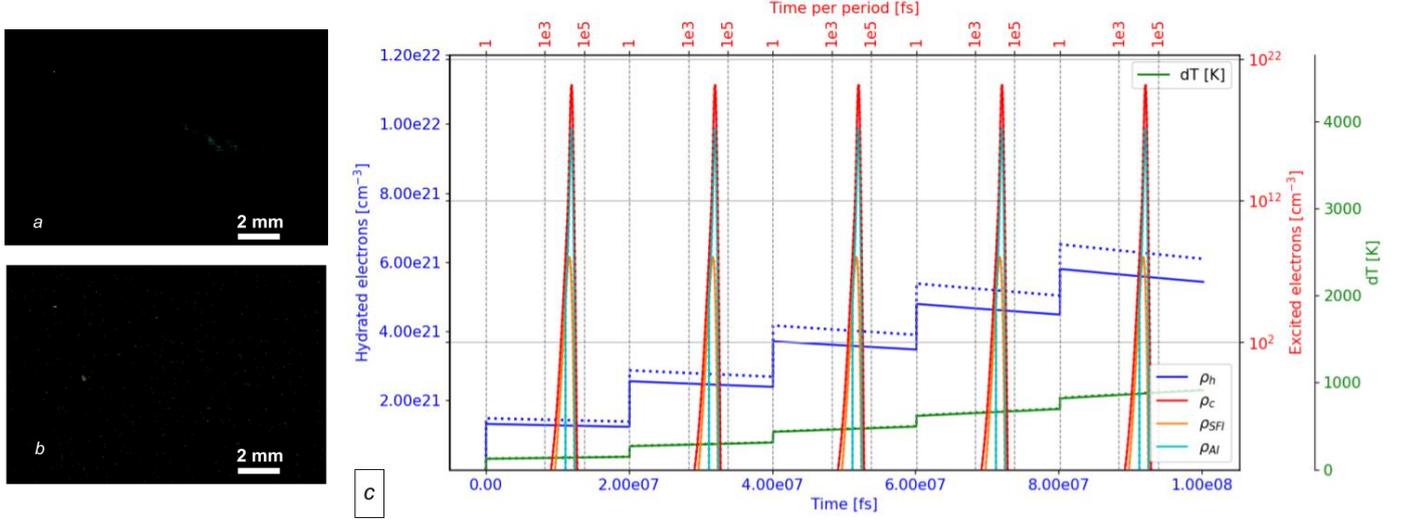

Fig. 8. Single pulses and burst water ablation ($\tau_p = 10$ ps, $F = 255$ mm): *a* - ablation ring in single distant pulses regime; *b* - ablation ring in burst regime with 5x pulses; *c* - theoretical evaluation of excited electron density dynamic and accumulation of the hydrated electrons (solid lines - $H_2O$, dotted lines - $D_2O$)

With a duration of 10 ps, the longest laser pulses showed no signs of intense ablation. A vanishing low glowing glow in the irradiated volume indicates that the laser is operating in a sub-threshold interaction regime with heavy water. Laser irradiation of light water has no detectable plasma glow, then the heavy water has it low too, but more detectable (1.0 and 2.6 a. u. respectively). Such a difference is in agreement with the higher absorption of laser beams in heavy water. However, even the burst regime with five laser pulses did not initiate an intense plasma glow in the optical spectrum for the longest laser pulses (Fig 2, *d*). On the other hand, the IR thermal signal from irradiated water volume is high enough (Fig. 3, *d*). The higher heating in the focal plane of the water highlights the considerable energy transfers from the laser beam to the surrounding liquid. This indicates that thermo-optical processes were indeed occurring, but the self-focusing of the laser beam did not reach the threshold needed to cause optical breakdown. In the corresponding calculations, the self-focusing was considered just below the threshold intensity, where the evaluated free electron density is lower than $\rho_{th}$. But even in such a sub-threshold regime the accumulation of hydrated electrons is high enough in comparison to the shortest laser pulses: from $1.32 \cdot 10^{21} - 1.48 \cdot 10^{21}$ cm$^{-3}$ up to $5.42 \cdot 10^{21} - 6.09 \cdot 10^{21}$ cm$^{-3}$ for the light and heavy water respectively.

*Irradiation with short focus length ($F = 100$ mm)*

For the wider focusing angle, laser beam filamentation is reduced, making it more challenging to achieve the power density required for optical breakdown [20]. Notably, only 1 ps laser pulses initiate detectable plasma glow with a significant accumulation effect. The important role of accumulation effects is highlighted by the increase in plasma glow intensity with increasing laser pulse numbers - from 8 a.u. for a single laser pulse up to 44 a.u. for 5 laser pulses in one burst (Fig 2, *c*). Such intensity of the single laser pulses is relatively low, but every next laser pulse in a burst leads to an increase of glow intensity. On the other hand, similar experiments with long focus give the highest intensity of plasma glow already with single pulses (compare $F = 100$ mm and $F = 255$ mm on Fig. 2). It means that in experiments with 100 mm focal length with ungrouped 1 ps laser pulses the critical density of excited electrons was not achieved, in contrast to 250 mm focal length experiments. Based on this, the self-focusing of the laser beam was set to a minimum required to achieve near-threshold electron density $\rho_c \sim 1.8 \cdot 10^{20}$ cm$^{-3}$. For a burst of 5 laser pulses with 1 ps duration each, this results in an accumulation of hydrated electrons exceeding $1.9 \cdot 10^{21}$ cm$^{-3}$. Other laser pulse durations did not reach the threshold density of free electrons and did not exhibit a significant accumulation effect in the burst (see Table 2, $F = 100$ mm). The observed interaction of the laser beam with the water volume under these conditions resembled the sub-threshold regime seen with $F = 255$ mm and 10 ps laser pulses (see Fig. 8, *a* and *b*). This sub-threshold regime ($\rho_c < 1.8 \cdot 10^{20}$ cm$^{-3}$) was used for calculations involving 240 fs and 10 ps laser pulse durations (Table 2, bottom right corner). The weaker self-focusing effect observed with the wider angle laser beam is corroborated by IR temperature measurements (Fig. 3, *c*). Due to the lower absorption of laser pulse energy in the focal volume, a larger portion of the energy penetrates into deeper water layers, leading to a more uniform distribution of water heating between the upper and lower layers (compare $F = 100$ mm in Fig. 3, *c* and $F = 255$ mm in Fig. 3, *d*).

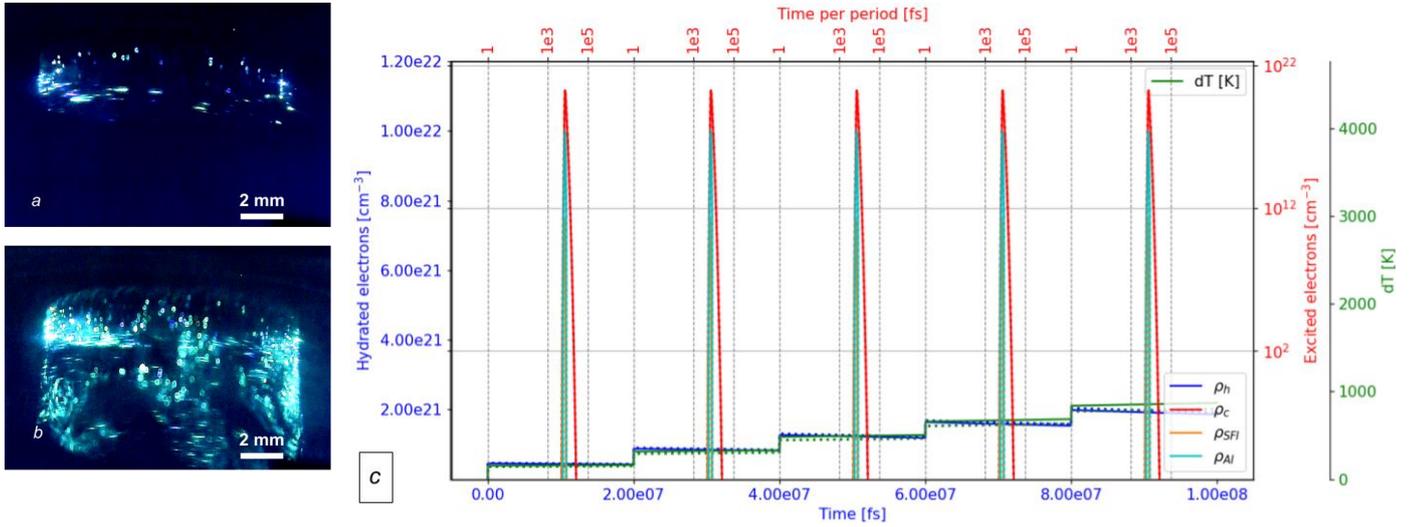

Fig. 9. Single pulses and burst water ablation ($\tau_p = 1$ ps, $F = 100$ mm): *a* - ablation ring in single distant pulses regime; *b* - ablation ring in burst regime with 5x pulses; *c* - theoretical evaluation of excited electron density dynamic and accumulation of the hydrated electrons (solid lines - $H_2O$, dotted lines - $D_2O$)

### 5.2. Thermal energy accumulation

For both regimes, distant single pulses and bursts, the delivered portion of the laser beam energy is the same. Thermocouple measurements confirmed that the amount of absorbed energy depended primarily on the focusing angle rather than on whether the laser was operated in single pulse or burst mode (Fig. 4). Table 3 presents the absorbed energy values for a single laser pulse in water at various focusing angles. The absorbed energy is calculated by dividing the total thermal energy in the water by the total number of laser pulses delivered during ring scanning. It is interesting to compare the experimental efficiency of the light absorption with the efficiency predicted by the first-order excitation model. In the biggest cases the predicted absorption is lower in comparison with experimental measurements. The exception to the above was the irradiation of water by 1 ps laser pulses, where the efficiency reached up to 20%. Such a difference can be explained by exceedingly high-intensity plasma glow, which bypasses laser pulse energy transfer into thermal energy, but directly converts into plasma plume light radiation. Of course, the ablation plasma plume explosion is not an equilibrium process, especially in the case of the burst regime. This is one of the reasons why the predicted thermal accumulation in the irradiated volume gives extremely high values of water temperature increasing (green line on Fig's. 6 - 9). Better temperature prediction can be achieved by including in calculations the thermal emission losses and fast explosion dynamic in the theoretical model. The theoretical predictions presented here can be used to evaluate relative thermal energy accumulation. This is particularly evident with 1 ps laser pulses, where the maximum heat accumulation is five times higher compared to other pulse durations.

Table 3. Comparison of absorbance efficiency measured by thermocouple and theoretically calculated

| Focusing (mm) | Pulse duration (fs) | Absorbed per pulse (μJ) (measured $H_2O$ \| $D_2O$) | Efficiency measured % (thermal measured $H_2O$ \| $D_2O$) | Absorbed on electrons (μJ) (modelling $H_2O$ \| $D_2O$) | Efficiency calculated % (modelling $H_2O$ \| $D_2O$) |
|---|---|---|---|---|---|
| 255 | 240 | 5.13 \| 3.04 | 12.83 \| 7.60 | 1.80 \| 1.64 | 4.50 \| 4.10 |
|  | 1000 | 5.91 \| 4.53 | 14.78 \| 11.33 | 8.09 \| 8.29 | 20.25 \| 20.73 |
|  | 10000 | 6.25 \| 2.7 | 15.63 \| 6.75 | 1.23 \| 1.33 | 3.08 \| 3.33 |
| 100 | 240 | 2.69 \| 3.51 | 6.73 \| 8.78 | 1.83 \| 1.58 | 4.58 \| 3.94 |
|  | 1000 | 3.39 \| 4.86 | 8.48 \| 12.15 | 1.69 \| 1.83 | 4.23 \| 4.58 |
|  | 10000 | 2.48 \| 3.35 | 6.20 \| 8.38 | 1.34 \| 1.46 | 3.35 \| 3.64 |

The theoretical predictions presented here can be used to assess relative thermal energy accumulation, especially for 1 ps laser pulses, where the maximum heat accumulation is five times higher compared to other pulse durations. It is important to note that, in the actual process of water irradiation, the laser beam absorbs energy along its entire path through the water volume, including after reflecting off the vessel bottom. This understanding of the absorption process is confirmed by measurements with the NIR sensor and IR camera, where higher water heating corresponds to lower detected scattering of the laser beam (see Fig. 3). The observed low efficiency of absorption in most predicted cases, except for 1 ps laser pulses, can be explained by the laser beam's extended path through the water, including reflections.

### 6. Discussion

In our experiments, the minimal time interval between laser pulses was 20 ns in the burst regime. The processes of excitation and recombination in the electron subsystem occur within a few picoseconds – hundreds of times shorter than the burst interval. This indicates that the excited state of electrons does not directly influence the subsequent laser pulse in a burst. Therefore, the observed step-by-step increase in plasma ablation glow in the burst regime cannot be attributed to interactions between laser pulses with conduction band electrons remaining from previous pulses. Instead, the rapid recombination of excited electrons leads to local heat accumulation within the focal volume, with maximum heat accumulation estimated to reach thousands of Kelvins. This is significantly higher than the critical temperature of 147.7 K determined by A. Vogel and co-authors [28]. Although the actual temperature in the irradiated volume is likely lower, theoretical calculations help estimate the relative heat accumulation during the burst regime. These calculations could be refined further by considering emission losses and fast explosion dynamics.

Our model also accounts for electrons trapped in hydrated states. After laser irradiation, the density of hydrated electrons remains high, around $10^{21}$ cm$^{-3}$, long after the laser pulse. This high density of hydrated electrons, combined with elevated temperatures, serves as a core for plasma formation with each subsequent laser pulse. Including hydrated electrons as an additional initial state density in the equation for maximum plasma intensity (Eq. 3) results in only minor changes in plasma formation – on the order of tenths. This outcome aligns with findings by Linz et al., who reported that the number of electrons excited via SFI is two orders of magnitude smaller than those excited through AI [3].

It should also be noted that hydrated electrons in water might form clusters, potentially affecting the physical properties of the surrounding water [31], [33], [39]. This state of overheated, electron-rich water likely modifies the optical properties of the irradiated volume [10], [40]–[42]. Other studies have experimentally shown that pressure changes in water containing hydrated electrons have a near-linear effect on the optical absorption of hot heavy water [43]. Although the exact mechanism by which hydrated electrons influence the optical properties of overheated water remains under discussion, this thermo-electron accumulation effect should not be overlooked. This effect was previously detected in studies of burst regime applications in light water [5], and we observed a similar accumulation effect in heavy water. This is evident in Fig. 2, *c* for 1 ps pulse duration with a 100 mm focus, and Fig. 2, *d* for 250 fs pulse duration with a 255 mm focus, where each successive laser pulse in a burst induces a step-by-step increase in plasma glow emission. For both focus distances, 1 ps laser pulses were found to be the most efficient. Interestingly, a similar optimal pulse duration of around 1 ps was also unexpectedly identified in our previous work on stainless steel ablation [44].

When examining the behavior of water under the shortest and longest laser pulses in the burst regime, it was observed that the accumulation of hydrated electrons is minimal during the shortest pulses. This is due to the limited interaction time between the laser pulse and water, which prevents the generation of large quantities of hydrated electrons, with intense electrons recombining and emitting light. While the longest pulses lead to significant hydration of electrons, their heat accumulation is similar to that of the shortest pulses (see Fig. 7 and Fig. 8). However, they do not produce a notable plasma glow (refer to Fig. 2, *d*). It is possible that the imbalance between hydrated electron accumulation and the relatively slow heat accumulation in the longest pulses negatively impacts plasma formation during the burst regime. This explanation aligns with the 1 ps experiments, where the rapid increase in hydrated electron density corresponds with a significant rise in accumulated temperature.

Additionally, a parity effect on ablation glow was observed in heavy water (Fig. 2, *d* – 240 fs for heavy water). A similar effect was noted in previous studies on the ablation of metallic targets in burst regimes [45], [46]. For solid-state targets, the parity effect was attributed to the interaction of secondary laser pulses with the shock wave front, which pushes back the ablation material. The third pulse then interacts with the overheated, suppressed ablation area [45]. In water ablation, a similar interaction between the secondary laser pulse and the ablation shock wave front likely occurs [5]. This parity effect is more pronounced in heavy water, possibly due to its higher density and lower sound speed compared to light water [47]. The interaction of ultrashort laser pulses burst with the rare state of plasma plume forming can be the most important factor for achieving higher energy density in the focal point. Such plasma energy accumulation plays an important role in the intensity of plasma glow, alongside the laser-induced effects on the irradiated area from pulse to pulse. The interaction of ultrashort laser pulse bursts with the rarefied state of the ablation plume is crucial for enhancing plasma energy. In burst mode, the irradiated area undergoes multiple interactions with successive laser pulses, which progressively alter its optical, electrical, chemical, and thermal properties due to the accumulation of hydrated electrons and thermal energy. This repeated interaction during the burst regime results in higher energy for excited electrons and ions compared to single-pulse ablation. In contrast, with a single pulse of equivalent total energy, the laser interacts with the equilibrium state of the water, and the plasma plume develops only after the pulse has ended.

The combination of thermo-electron accumulation and parity effects initiated in the burst regime presents a promising approach for achieving more efficient energy transfer from laser pulses to the ablation volume. The ability to rapidly alter the water state during burst irradiation can be harnessed for developing hierarchical structures in underwater solid texturing and for intensifying fusion effects in the presence of a solid-state wall [48], [49]. Our future work will focus on investigating the interaction of GHz bursts at interfaces between different media and the influence of varying pulse-to-pulse duty cycles on thermo-electron accumulation effects.

**Conclusion**

The burst laser ablation in both light and heavy water was systematically studied using short-focus and long-focus objectives. The most significant finding is the pronounced effect of the burst regime on plasma glow intensity. Specifically, it was observed that the burst mode produced plasma glow intensities more than 18 times higher than those generated by single pulses at the same power density, in both light and heavy water.

Additionally, an unexpected effect was discovered during ablation with a long-focus objective: lower power density laser pulses ($2.6 \cdot 10^{13}$ W/cm$^2$) produced plasma glow intensities four times higher than those from higher power density pulses ($8 \cdot 10^{13}$ W/cm$^2$). This result may be attributed to the decreased optical threshold energy associated with smaller focusing angles, combined with the increased ablation threshold for shorter laser pulses. Light water demonstrated a greater sensitivity to absorption changes based on laser beam focusing compared to heavy water.

In both short-focus and long-focus experiments, the burst regime significantly reduced the ablation threshold, particularly when using 1 picosecond laser pulses, which proved to be the most effective. The combination of optimal pulse duration, reduced ablation thresholds in burst mode, and a narrow laser beam holds promise for future experiments aimed at improving DD-neutron fusion efficiency.

Theoretical calculations suggest that the observed increase in pulse-to-pulse plasma glow in the burst regime cannot be solely explained by the interaction of successive laser pulses with excited electrons in the conduction band. A more plausible explanation lies in the thermo-electron accumulation of absorbed thermal energy, facilitated by the presence of high-density hydrated electrons. The direct impact of this laser-modified state of water warrants further investigation. One potential avenue for enhancing the efficiency of laser water ablation in burst mode could involve optimizing the pulse-to-pulse duty cycle in the presence of a solid wall interface.

**Declaration of generative AI and AI-assisted technologies in the writing process**

During the preparation of this work the authors used ChatGPT in order to enhance grammar and style. After using this tool/service, the authors reviewed and edited the content as needed and take full responsibility for the content of the published article.

**Acknowledgments**

The work was supported by the Ministry of Education, Youth and Sports of the Czech Republic (OP JAK program, MEBioSys project, No. CZ.02.01.01/00/22_008/0004634, co-funded by EU).

**References**

[1] P. K. Kennedy, "A First-Order Model for Computation of Laser-Induced Breakdown Thresholds in Ocular and Aqueous Media: Part I—Theory," *IEEE J. Quantum Electron.*, vol. 31, no. 12, pp. 2241–2249, 1995, doi: 10.1109/3.477753.

[2] L. V. Keldysh, "Ionization in the Field of a Strong Electromagnetic Wave," *Sov. Phys. JETP*, vol. 20, no. 5, pp. 1307–1314, 1965.

[3] N. Linz, S. Freidank, X. X. Liang, and A. Vogel, "Wavelength dependence of femtosecond laser-induced breakdown in water and implications for laser surgery," *Phys. Rev. B*, vol. 94, no. 2, 2016, doi: 10.1103/PhysRevB.94.024113.

[4] A. Vogel, J. Noack, G. Hüttman, and G. Paltauf, "Mechanisms of femtosecond laser nanosurgery of cells and tissues," *Appl. Phys. B Lasers Opt.*, vol. 81, no. 8, pp. 1015–1047, 2005, doi: 10.1007/s00340-005-2036-6.

[5] Z. Qian, A. Covarrubias, A. W. Grindal, M. K. Akens, L. Lilge, and R. S. Marjoribanks, "Dynamic absorption and scattering of water and hydrogel during high-repetition-rate (>100 MHz) burst-mode ultrafast-pulse laser ablation," *Biomed. Opt. Express*, vol. 7, no. 6, p. 2331, Jun. 2016, doi: 10.1364/BOE.7.002331.

[6] R. S. Marjoribanks *et al.*, "Ablation and thermal effects in treatment of hard and soft materials and biotissues using ultrafast-laser pulse-train bursts," *Photonics Lasers Med.*, vol. 1, no. 3, pp. 155–169, 2012, doi: 10.1515/plm-2012-0020.

[7] A. Žemaitis, M. Gaidys, M. Brikas, P. Gečys, G. Račiukaitis, and M. Gedvilas, "Advanced laser scanning for highly-efficient ablation and ultrafast surface structuring: experiment and model," *Sci. Rep.*, vol. 8, no. 1, p. 17376, Dec. 2018, doi: 10.1038/s41598-018-35604-z.

[8] N. Sakakibara, T. Ito, K. Terashima, Y. Hakuta, and E. Miura, "Dynamics of solvated electrons during femtosecond laser-induced plasma generation in water," *Phys. Rev. E*, vol. 102, no. 5, p. 053207, Nov. 2020, doi: 10.1103/PhysRevE.102.053207.

[9] J. M. Wiesenfeld and E. P. Ippen, "Dynamics of electron solvation in liquid water," *Chem. Phys. Lett.*, vol. 73, no. 1, pp. 47–50, 1980, doi: 10.1016/0009-2614(80)85199-2.

[10] D. N. Nikogosyan, A. A. Oraevsky, and V. I. Rupasov, "Two-photon ionization and dissociation of liquid water by powerful laser UV radiation," *Chem. Phys.*, vol. 77, no. 1, pp. 131–143, May 1983, doi: 10.1016/0301-0104(83)85070-8.

[11] J. Tang and Z. Wang, "Hydrated Electron Dynamics and Stimulated Raman Scattering in Water Induced by Ultrashort Laser Pulses," *Molecules*, vol. 29, no. 6, p. 1245, Mar. 2024, doi: 10.3390/molecules29061245.

[12] M. Spellauge *et al.*, "Photomechanical Laser Fragmentation of IrO2 Microparticles for the Synthesis of Active and Redox-Sensitive Colloidal Nanoclusters," *Small*, vol. 19, no. 10, pp. 1–13, 2023, doi: 10.1002/smll.202206485.

[13] M. Spellauge, C. Doñate-Buendía, S. Barcikowski, B. Gökce, and H. P. Huber, "Comparison of ultrashort pulse ablation of gold in air and water by time-resolved experiments," *Light Sci. Appl.*, vol. 11, no. 1, 2022, doi: 10.1038/s41377-022-00751-6.

[14] J. Hah *et al.*, "High repetition-rate neutron generation by several-mJ, 35 fs pulses interacting with free-flowing D2O," *Appl. Phys. Lett.*, vol. 109, no. 14, 2016, doi: 10.1063/1.4963819.

[15] J. Krása and D. Klír, "Scaling of Laser Fusion Experiments for DD-Neutron Yield," *Front. Phys.*, vol. 8, no. September, pp. 3–9, 2020, doi: 10.3389/fphy.2020.00310.

[16] J. Alvarez *et al.*, "Laser Driven Neutron Sources: Characteristics, Applications and Prospects," *Phys. Procedia*, vol. 60, no. C, pp. 29–38, 2014, doi: 10.1016/j.phpro.2014.11.006.

[17] N. Tinne, B. Kaune, A. Krüger, and T. Ripken, "Interaction mechanisms of cavitation bubbles induced by spatially and temporally separated fs-laser pulses," *PLoS One*, vol. 9, no. 12, 2014, doi: 10.1371/journal.pone.0114437.

[18] D. Hakoume, L. A. Dombrovsky, D. Delaunay, and B. Rousseau, "An experimental determination of near-infrared properties of polypropylene and composite material containing polypropylene and glass fibers," *16th Eur. Conf. Compos. Mater. ECCM 2014*, no. July, 2014.

[19] D. Bergstrom *et al.*, "Noise properties of a corner-cube Michelson interferometer LWIR hyperspectral imager," in *Infrared Technology and Applications XXXVI*, B. F. Andresen, G. F. Fulop, and P. R. Norton, Eds., Apr. 2010, p. 76602F. doi: 10.1117/12.851433.

[20] A. Vogel, K. Nahen, D. Theisen, and J. Noack, "Plasma formation in water by picosecond and nanosecond Nd:YAG laser pulses. I. Optical breakdown at threshold and superthreshold irradiance," *IEEE J. Sel. Top. Quantum Electron.*, vol. 2, no. 4, pp. 847–860, Dec. 1996, doi: 10.1109/2944.577307.

[21] G. Sinibaldi *et al.*, "Laser induced cavitation: Plasma generation and breakdown shockwave," *Phys. Fluids*, vol. 31, no. 10, 2019, doi: 10.1063/1.5119794.

[22] A. Aguilar *et al.*, "Ultrafast laser induced cavitation bubbles in water in the presence of optical aberrations," *Opt. InfoBase Conf. Pap.*, vol. 860, no. 1996, p. 6654, 2021.

[23] J. G. Bayly, V. B. Kartha, and W. H. Stevens, "The absorption spectra of liquid phase H2O, HDO and D2O from 0·7 µm to 10 µm," *Infrared Phys.*, vol. 3, no. 4, pp. 211–222, Dec. 1963, doi: 10.1016/0020-0891(63)90026-5.

[24] Y. R. Shen, "The principles of nonlinear optics," New York: Wiley, 1985, pp. 528–539.

[25] J. Noack and A. Vogel, "Laser-induced plasma formation in water at nanosecond to femtosecond time scales: calculation of thresholds, absorption coefficients, and energy density," *IEEE J. Quantum Electron.*, vol. 35, no. 8, pp. 1156–1167, 1999, doi: 10.1109/3.777215.

[26] Z. Yang, C. Zhang, H. Zhang, and J. Lu, "Transient electron temperature and density changes in water breakdown induced by femtosecond laser pulses," *Opt. Commun.*, vol. 546, no. July, p. 129803, 2023, doi: 10.1016/j.optcom.2023.129803.

[27] X.-X. Liang, Z. Zhang, and A. Vogel, "Multi-rate-equation modeling of the energy spectrum of laser-induced conduction band electrons in water," *Opt. Express*, vol. 27, no. 4, p. 4672, 2019, doi: 10.1364/oe.27.004672.

[28] A. Vogel, N. Linz, S. Freidank, and G. Paltauf, "Femtosecond-laser-induced nanocavitation in water: Implications for optical breakdown threshold and cell surgery," *Phys. Rev. Lett.*, vol. 100, no. 3, 2008, doi: 10.1103/PhysRevLett.100.038102.

[29] E. S. Efimenko, Y. A. Malkov, A. A. Murzanev, and A. N. Stepanov, "Femtosecond laser pulse-induced breakdown of a single water microdroplet," *J. Opt. Soc. Am. B*, vol. 31, no. 3, p. 534, 2014, doi: 10.1364/josab.31.000534.

[30] F. Uhlig, O. Marsalek, and P. Jungwirth, "Electron at the Surface of Water: Dehydrated or Not?," *J. Phys. Chem. Lett.*, vol. 4, no. 2, pp. 338–343, Jan. 2013, doi: 10.1021/jz3020953.

[31] F. Uhlig and P. Jungwirth, "Embedded Cluster Models for Reactivity of the Hydrated Electron," *Zeitschrift für Phys. Chemie*, vol. 227, no. 9–11, pp. 1583–1593, Jan. 2013, doi: 10.1524/zpch.2013.0402.

[32] J. M. Herbert and M. P. Coons, "The Hydrated Electron," *Annu. Rev. Phys. Chem.*, vol. 68, no. 1, pp. 447–472, May 2017, doi: 10.1146/annurev-physchem-052516-050816.

[33] D. H. Paik, I.-R. Lee, D.-S. Yang, J. S. Baskin, and A. H. Zewail, "Electrons in Finite-Sized Water Cavities: Hydration Dynamics Observed in Real Time," *Science (80-. ).*, vol. 306, no. 5696, pp. 672–675, Oct. 2004, doi: 10.1126/science.1102827.

[34] B. F. Bachman, D. Zhu, J. Bandy, L. Zhang, and R. J. Hamers, "Detection of Aqueous Solvated Electrons Produced by Photoemission from Solids Using Transient Absorption Measurements," *ACS Meas. Sci. Au*, vol. 2, no. 1, pp. 46–56, Feb. 2022, doi: 10.1021/acsmeasuresciau.1c00025.

[35] D. Moskal, J. Martan, and M. Kucera, "Shifted Laser Surface Texturing (sLST) in Burst Regime," *J. Laser Micro/Nanoengineering*, vol. 14, no. 2, pp. 179–185, Sep. 2019, doi: 10.2961/jlmn.2019.02.0011.

[36] F. Bauer, A. Michalowski, T. Kiedrowski, and S. Nolte, "Heat accumulation in ultra-short pulsed scanning laser ablation of metals," *Opt. Express*, vol. 23, no. 2, p. 1035, Jan. 2015, doi: 10.1364/OE.23.001035.

[37] J. Martan, L. Prokešová, D. Moskal, B. C. Ferreira de Faria, M. Honner, and V. Lang, "Heat accumulation temperature measurement in ultrashort pulse laser micromachiningHeat accumulation temperature measurement in ultrashort pulse laser micromachining," *Int. J. Heat Mass Transf.*, vol. 168, p. 120866, 2021, doi: 10.1016/j.ijheatmasstransfer.2020.120866.

[38] A. COUAIRON and A. MYSYROWICZ, "Femtosecond filamentation in transparent media," *Phys. Rep.*, vol. 441, no. 2–4, pp. 47–189, Mar. 2007, doi: 10.1016/j.physrep.2006.12.005.

[39] K. S. Kim *et al.*, "The Nature of a Wet Electron," *Phys. Rev. Lett.*, vol. 76, no. 6, pp. 956–959, Feb. 1996, doi: 10.1103/PhysRevLett.76.956.

[40] D. I. Kovsh, D. J. Hagan, and E. W. Van Stryland, "Numerical modeling of thermal refraction in liquids in the transient

regime," *Opt. Express*, vol. 4, no. 8, p. 315, Apr. 1999, doi: 10.1364/OE.4.000315.
[41] D. M. Bartels, K. Takahashi, J. A. Cline, T. W. Marin, and C. D. Jonah, "Pulse Radiolysis of Supercritical Water. 3. Spectrum and Thermodynamics of the Hydrated Electron," *J. Phys. Chem. A*, vol. 109, no. 7, pp. 1299–1307, Feb. 2005, doi: 10.1021/jp0457141.
[42] S. R. J. Brueck, H. Kildal, and L. J. Belanger, "Photo-acoustic and photo-refractive detection of small absorptions in liquids," *Opt. Commun.*, vol. 34, no. 2, pp. 199–204, Aug. 1980, doi: 10.1016/0030-4018(80)90014-0.
[43] J.-P. Jay-Gerin, M. Lin, Y. Katsumura, H. He, Y. Muroya, and J. Meesungnoen, "Effect of water density on the absorption maximum of hydrated electrons in sub- and supercritical water up to 400 °C," *J. Chem. Phys.*, vol. 129, no. 11, Sep. 2008, doi: 10.1063/1.2978955.
[44] D. Moskal, J. Martan, M. Honner, C. Beltrami, M.-J. Kleefoot, and V. Lang, "Inverse dependence of heat accumulation on pulse duration in laser surface processing with ultrashort pulses," *Int. J. Heat Mass Transf.*, vol. 213, p. 124328, Oct. 2023, doi: 10.1016/j.ijheatmasstransfer.2023.124328.
[45] D. J. Förster *et al.*, "Shielding effects and re-deposition of material during processing of metals with bursts of ultra-short laser pulses," *Appl. Surf. Sci.*, vol. 440, pp. 926–931, May 2018, doi: 10.1016/j.apsusc.2018.01.297.
[46] A. Žemaitis, P. Gečys, M. Barkauskas, G. Račiukaitis, and M. Gedvilas, "Highly-efficient laser ablation of copper by bursts of ultrashort tuneable (fs-ps) pulses," *Sci. Rep.*, vol. 9, no. 1, p. 12280, Aug. 2019, doi: 10.1038/s41598-019-48779-w.
[47] S. Lago, P. A. Giuliano Albo, and G. Cavuoto, "Speed of sound measurements in deuterium oxide (D2O) at temperatures between (276.97 and 363.15) K and at pressures up to 210 MPa," *Fluid Phase Equilib.*, vol. 506, p. 112401, Feb. 2020, doi: 10.1016/j.fluid.2019.112401.
[48] D. Zhang, B. Ranjan, T. Tanaka, and K. Sugioka, "Multiscale Hierarchical Micro/Nanostructures Created by Femtosecond Laser Ablation in Liquids for Polarization-Dependent Broadband Antireflection," *Nanomaterials*, vol. 10, no. 8, p. 1573, Aug. 2020, doi: 10.3390/nano10081573.
[49] R. V. Volkov *et al.*, "Neutron generation in dense femtosecond laser plasma of a structured solid target," *J. Exp. Theor. Phys. Lett.*, vol. 72, no. 8, pp. 401–404, Oct. 2000, doi: 10.1134/1.1335116.